\documentclass[10pt]{article}

\usepackage{url}
\usepackage{amsmath}
\usepackage{amssymb}
\usepackage{natbib}
\usepackage{graphicx}
\newcommand{\Nc}{\mathcal{N}} % calligraphic N
\renewcommand{\vec}[1]{\mathbf{#1}}

\newcommand{\vecop}{\operatorname{vec}} % calligraphic N

\begin{document}

\title{Sparse Causal Discovery \\in Multivariate Time Series}

\author{Stefan Haufe\footnote{haufe@cs.tu-berlin.de}\\Berlin Institute of Technology \and Guido Nolte\\Fraunhofer First, Berlin
       \and Klaus-Robert M\"uller \\Berlin Institute of Technology \and
        Nicole Kr\"amer \\Berlin Institute of Technology
       }

\maketitle

\begin{abstract}%   <- trailing '%' for backward compatibility of .sty file
Our goal is to estimate causal interactions in multivariate time series. Using vector autoregressive (VAR) models, these can be defined based on non-vanishing coefficients belonging to respective time-lagged instances. As in most cases a parsimonious causality structure is assumed, a promising approach to causal discovery consists in fitting VAR models with an additional sparsity-promoting regularization. Along this line we here propose that sparsity should be enforced for the subgroups of coefficients that belong to each pair of time series, as the absence of a causal relation requires the coefficients for all time-lags to become jointly zero. Such behavior can be achieved by means of $\ell_{1,2}$-norm regularized regression, for which an efficient active set solver has been proposed recently. Our method is shown to outperform standard methods in recovering simulated causality graphs. The results are on par with a second novel approach which uses multiple statistical testing.
\end{abstract}


\section{Introduction}
\label{sec:introduction}
Causality is commonly defined based on the widely accepted assumption that an effect is always preceded by its cause. \cite{granger} postulates a measure of causal influence between two time series (\emph{Granger Causality}). In a nutshell, time series $z_i$ Granger-causes time series $z_j$ if knowledge of past values of $z_i$ improves the prediction of $z_j$ (compared to only using past values of $z_j$). In the case of a set $F=\{z_1,\ldots, z_M\}$ of time series, the pairwise analysis may lead to spurious detection of a causal relation. For this reason it is advisable to include the set $F\setminus \{z_i,z_j\}$ of all additional observable time series in both prediction tasks. Note that this approach resolves the problem of common hidden factors $z_*$ if $z_* \in F$ . (If common factors are not observable, Granger causality fails and we refer to  \cite{Nolte0801} for a detailed discussion and a remedy in form of the Phase Slope Index.) While this approach, to which we refer as \emph{complete} Granger Causality, is practical, a more elegant way to deal with multivariate data is to handle all potential causal relations between all time series at once. In this paper, we assume a linear dynamics of the underlying system, which leads to the vector autoregressive (VAR) model.

In many applications the true causality graph is assumed to be sparse, i.e. only a few causal interactions between time series are expected. However, both Ordinary Least Squares (OLS) and Ridge Regression, which are usually used for fitting VAR models, are known for producing dense coefficients. Only recently \cite{Valdes0501} have proposed to enforce estimation of sparse AR coefficients using $\ell_1$-norm regularized models such as the Lasso \citep{lasso}.

In this paper we propose a novel sparse approach which -- unlike Lasso -- accounts for the fact that the absence of a causal relation between $z_i$ and $z_j$ requires all AR coefficients belonging to that certain pair of time series to be jointly zero. Furthermore, we consider Ridge Regression in combination with the multiple statistical testing procedure provided by \cite{Hothorn0801}. More details on the methodology are given in Section \ref{sec:methods}. These methods are  evaluated and compared to standard approaches in extensive simulations.
%----------------------
\section{Background}
\label{sec:background}
%----------------------
In this section, we briefly summarize related approaches to estimate sparse vector autoregressive models in the context of causal discovery. We roughly distinguish between sparse estimation methods and testing strategies.

Given a multivariate time series $\vec{z}(t) \in \mathbb{R}^M$, a linear vector autoregressive process of order $P$ is defined as

\begin{eqnarray}
\label{eq:varmodel}
\vec{z}(t) & = & \sum_{p=1} ^P A^{(p)} \vec{z}({t-p}) + \boldsymbol{\varepsilon}(t)\,,
\end{eqnarray}
where $A^{(p)} \in \mathbb{R}^{M \times M}$, $\boldsymbol{\varepsilon} \sim {\cal N}(\vec{0}, \sigma^2 I)$ and $t \in \mathbb{Z}$ indicates time. Hence, the signal at time $t$ is modeled as a linear combination of its $P$ past values and Gaussian measurement noise.  Inspired by the initial assumption that the cause should always precede the effect, we suggest the following definition of causality in the case of vector autoregressive models. We say that time series $z_i$ has a causal influence on time series $z_j$ if for at least one  $p \in \{ 1, \hdots, P \}$, the coefficient  $A^{(p)}_{ji}$ corresponding to the interaction between $z_j$ and $z_i$ at the $p$th time-lag is nonzero.

Thus, causal inference may be conducted by estimating the matrices $A^{(p)}$ from a sample $Z = \left(\vec{z}(1), \hdots, \vec{z}(T) \right)$. Let us introduce the following shortcuts. We denote by $A = \left( A^{(1)}, \hdots, A^{(P)} \right)^\top $ the matrix of all VAR coefficients and set  $X = \left( Z_1, \hdots, Z_P \right)$, $Y = Z_0$, $Z_p = \left( \vec{z}(P+1-p), \hdots, \vec{z}(T-p) \right)^\top $. Here  $\vecop( \cdot )$ denotes the vectorization operation.

\subsection{Sparsity}
Probably the most straightforward way to estimate a sparse VAR is to use $\ell_1$-regularization on the set of coefficients,
\begin{equation*}
 {\widehat A}^{\mbox{lasso}} = \arg \min_{A} \left\| \vecop (XA - Y) \right\|_2^2 + \lambda \left\| \vecop (A) \right\|_1  \;, \;\lambda \geq 0 \;.
\end{equation*}

Recently, \cite{Valdes0501} proposed a combination of VAR-estimation and the Lasso \citep{lasso}. While \cite{Valdes0501} only consider a VAR model of order $1$, there have been extensions to higher orders \citep[e.g.][]{Arnold0701}. However, we note in the latter case, Lasso is not used  on the VAR coefficients directly, but that the problem is transformed into the task of estimating partial correlation coefficients between time-lagged copies of the time series \citep[see also][]{Opgen0701}.

\subsection{Testing}
Just as in the case of sparse methods, it is often suggested to transform the regression task into the estimation of the matrix of partial correlation coefficients  between time-lagged copies of the time series. While  \cite{Drton0801} estimate the correlation matrix in an unregularized way, \cite{Opgen0701} propose a shrinkage estimator, which is  superior in the case of high-dimensional data \citep{Schaefer0501}. Afterwards, significant partial correlations are detected by controlling false discovery rates. While the latter approach is only tested for $P=1$, it is straightforward to extend it to higher order VAR's.
%------------------------------------
\section{Our Approach}\label{sec:methods}
%------------------------------------
In the following, we provide the details regarding the groupwise sparsity and the alternative testing strategy respectively.
\subsection{Ridge Regression and Multiple Testing} \label{subsec:ridge}
Under the assumption of Gaussian white noise it is natural to estimate the AR coefficients using regularized least squares, and probably the most straightforward way to do so is to use Ridge Regression,
\begin{equation}\label{eq:ridge}
  {\widehat A}^{\mbox{ridge}} = \arg \min_{A} \left\| \vecop (XA - Y) \right\|_2^2 + \lambda \left\| \vecop (A) \right\|_2^2 = (X^\top X + \lambda I)^{-1}X^\top Y \;, \lambda \geq 0 \;.
\end{equation}
Thanks to the Ridge penalty, Eq.~\eqref{eq:ridge} delivers solutions with small coefficients, which, however, are in general never exactly zero. In the strict sense of Granger, this corresponds to a fully-connected dependency graph, rendering Ridge Regression an improper candidate for sparse causal recovery. On the other side, many of the estimated coefficients are expected to be non-significant. Hence, we propose a sparsification through statistical testing. In contrast to  to e.g. bootstrapping, we derive $p$-values explicitly using the approximate distribution of the coefficients.

It is apparant from Eq.~\eqref{eq:ridge} that the estimation can be done independently for each column of $A$, and so does the testing. Let therefore $\boldsymbol{\alpha}_k$ denote the $k$th column of $A$ and let $\vec{y}_k = \left( z_k(P+1), \hdots, z_k(T) \right)^\top $. Neglecting the dependency of $X$ and $Y$, the Ridge coefficients depend linearly on $Y$, and we can conclude that under the null-hypothesis $H_0: \boldsymbol{\alpha}_k = 0$, we have $\widehat{\boldsymbol{\alpha}}_k \sim \Nc (\vec{0}, \sigma_k^2 \Sigma)$ with
\begin{eqnarray*}
\Sigma &=& \left( X^\top  X + \lambda I\right)^{-1} X^\top  X \left( X^\top  X + \lambda I\right)^{-1}\,.
\end{eqnarray*}
Furthermore, setting $H = X \left( X^\top  X + \lambda I \right) ^{-1} X^\top $ an estimate of the model variance $\sigma_k^2$ is given by
\begin{equation}\label{eq:sigma}
\widehat \sigma_k^2 = \frac{\left\| \vec{y}_k - H \vec{y}_k \right\|^2}{\operatorname{trace}\left( (I-H)(I-H^\top )\right)} \;.
\end{equation}
Using Eq.~\eqref{eq:sigma} we can now construct normalized test statistics $\widetilde \alpha_{ik} = \widehat \alpha_{ik} / \sqrt{ \sigma_k^2 \Sigma_{ii}}$ which are jointly normally distributed with $\widetilde{ \boldsymbol{\alpha}} \sim \Nc (\vec{0}, R)$ and $R_{ij} := \Sigma_{ij} / \sqrt{\Sigma_{ii} \Sigma_{jj}}$. Suppose we want to test all individual hypotheses $H_{0,i}:\alpha_{ik} = 0$ simultaneously, then, according to \cite{Hothorn0801}, the adjusted p-values are $p_i = 1 - g \left(R, |\widetilde \alpha_{ik}|\right)$. We reject a hypothesis, if the $p$-value is below the predefined significance level $\gamma$. Here,
\begin{equation}\label{eq:rej}
g(R,t) = P\left( \max_{i} \left|\widetilde \alpha_{ik} \right| \leq t\right) = \int\nolimits_{-t} ^t  \ldots \int\nolimits_{-t} ^t  \phi(\alpha_1,\ldots, \alpha_{MP}) \mbox{d}\alpha_1 \cdots \mbox{d}\alpha_{MP}
\end{equation}
and $\phi(\boldsymbol{\alpha})$ is the density function of the multivariate normal distribution $\Nc (\vec{0}, R)$.
%---------------------------------------------------
\subsection{Group Lasso}\label{sec:lasso}
%---------------------------------------------------
Sparse causal discovery using Ridge Regression is a two-step procedure and may possibly suffer from the aggregation of assumptions that enter in each step. Direct estimation of sparse VAR coefficients (e.g. via Lasso) is therefore desirable, as this would allow omission of the multiple significance testing step.  However, for higher order models, this approach is prone to selecting a different set of causal interactions for each of the $P$ time lags. We here suggest that this behavior can be overcome by enforcing \emph{joint sparsity} of the coefficient vectors that belong to a certain pair of time series. This corresponds to incorporating the prior belief that causal influences between time series are not restricted to only one particular time lag into the estimation. The positive effect of such modeling can be verified in Figure~\ref{fig2} (see Section \ref{sec:experiments} for more details).

The idea of imposing groupwise sparse coefficients leads to $\ell_{1, 2}$-norm regularized regression also known as the \emph{Group Lasso} \citep{Yuan0601,Meier0801}, which has also applications in Multiple Kernel Learning \citep{bachmkl,Sonnenburg0601} and the EEG/MEG inverse problem \citep[e.g.][]{Haufe0801}. The term $\ell_{1, 2}$-norm corresponds to an $\ell_{1}$-norm of a vector of $\ell_{2}$-norms. Our proposed objective is given by
\begin{eqnarray}\label{eq:glasso1}
  {\widehat A}^{\mbox{glasso}} &=& \arg \min_{A} \left\| \vecop (XA - Y) \right\|_2^2 \\
 \label{eq:glasso2}  \mbox{s.t.} && \left\| \left( A_{11}^{(1)}, \hdots, A_{MM}^{(P)} \right) \right\|_2 + \sum_{i \neq j} \left\| \left( A_{ij}^{(1)}, \hdots, A_{ij}^{(P)} \right) \right\|_2 \leq \kappa \;,
\end{eqnarray}
This penalty leads to a groupwise variable selection, i.e. a whole block of coefficients is jointly zero. Note that the first term in Eq.~\eqref{eq:glasso2} penalizes all $MP$ coefficients describing univariate relations. In this way, those coefficients are shrunk and hence, overfitting is avoided. Furthermore, we remark that it is also conceivable to to split the estimation of $A$ into $M$ subproblems (as suggested in Subsection~\ref{subsec:ridge}), which is desirable in large-scale scenarios.

Eqs.~\eqref{eq:glasso1} and  \eqref{eq:glasso2} define a non-differentiable but convex optimization problem which can be solved by means of Second-order Cone Programming (SOCP). For problems with sparse expected structure, however, the optimization can be carried out much more efficiently using the results of \cite{Roth0801}. By keeping a set of active coefficient groups, their algorithm needs to call the SOCP solver only for problem sizes far smaller than the original problem  -- leading to a considerable reduction of memory usage and computation time. In the experiments, we employ the active-set algorithm of \cite{Roth0801} in combination with a freely available SOCP solver \citep{sedumi}.
%-------------------------------------------
\section{Simulations}\label{sec:experiments}
%-------------------------------------------
We conduct a series of experiments in which the causal structure of simulated data has to be recovered. We compare the Group Lasso, standard Lasso, Ridge Regression with multiple testing and complete Granger Causality based on AR models. All four approaches are applied both with and without knowledge of the true model order. In the latter case $P=10$ is chosen for the reconstruction. For all methods considered, it is also possible to estimate the model order $P$, e.g. via cross-validation.

%-----------------------------------
\subsection{Setup} \label{sec:setup}
%-----------------------------------
 Each simulated data set consists of a multivariate time series with parameters $M=7$ and $T=1000$ that is generated by a random VAR process of order $P=5$ according to \eqref{eq:varmodel}. The distribution of the noise component $\boldsymbol{\varepsilon}(t)$ is chosen to be the standard normal distribution. The VAR coefficients for all but 10 randomly chosen pairs of time series are set to zero, yielding exactly 10 causal interactions. The non-zero coefficients are drawn randomly from ${\cal N} (0, 0.04 I)$. Each set of VAR coefficients is tested for the stability of its induced dynamical system by looking at the eigenvalues of the corresponding transition matrix. Only coefficients leading to stable systems (i.e those with transition matrices with eigenvalues of at most 1) are accepted. We consider the following three types of problems, for each of which we create 10 instances: 1) no noise is added to the data generated by the VAR model 2) the data is superimposed by Gaussian noise of approximately the same strength, which is uncorrelated (white) both across time and sensors 3) the data is superimposed by mixed noise of approximately the same strength, which is generated as a random instantaneous mixture of $M$ univariate AR processes of order 20. Note that in none of these cases the noise itself possesses a causal structure which would superimpose the true structure.

For measuring performance we consider Receiver Operating Characteristics (ROC) curves, which allow objective assessment of the performance in different regimes (e.g. very few false positives). As an additional measure of absolute performance we calculate the Area Under Curve (AUC). ROC curves and AUC values are averaged across the 10 problem instances and standard errors are computed for AUC.

Granger Causality is calculated using the Levinson-Wiggens-Robinson algorithm for fitting AR models \citep{marple87}, which is available in the open source Biosig toolbox \citep{biosig}. Note that for this particular method, we use Granger's original definition of causal influence instead of our coefficient-based approach. That is, for a pair of time series $z_i$ and $z_j$  we calculate the logarithm of the ratio of the residuals of the two AR models 1) including interactions and 2) excluding interactions between $z_i$ and $z_j$ (\emph{Granger score}). This score is divided by its standard deviations as estimated by the jackknife. To obtain a ROC-curve, the Granger score is threshold at different values, ranging from completely sparse to completely dense solutions.

For Ridge Regression, the regularization parameter $\lambda$ is chosen via $10$-fold cross-validation (with respect to prediction accuracy). For this value of $\lambda$, we derive the test statistics defined in Subsection\ref{subsec:ridge}. The multidimensional integrals in Eq.~\eqref{eq:rej} are computed using Monte Carlo sampling according to \cite{Genz92}.  ROC-curves are constructed by
varying the significance level $\gamma$.

For Lasso and Group Lasso, solutions ranging from completely sparse to completely dense are obtained through variation of the regularizing constant $\lambda$ and $\kappa$ respectively.

\subsection{Results and Discussion}
\begin{figure}[htb]
\begin{center}
\begin{tabular}{ccccc}
\includegraphics[width=0.17\textwidth]{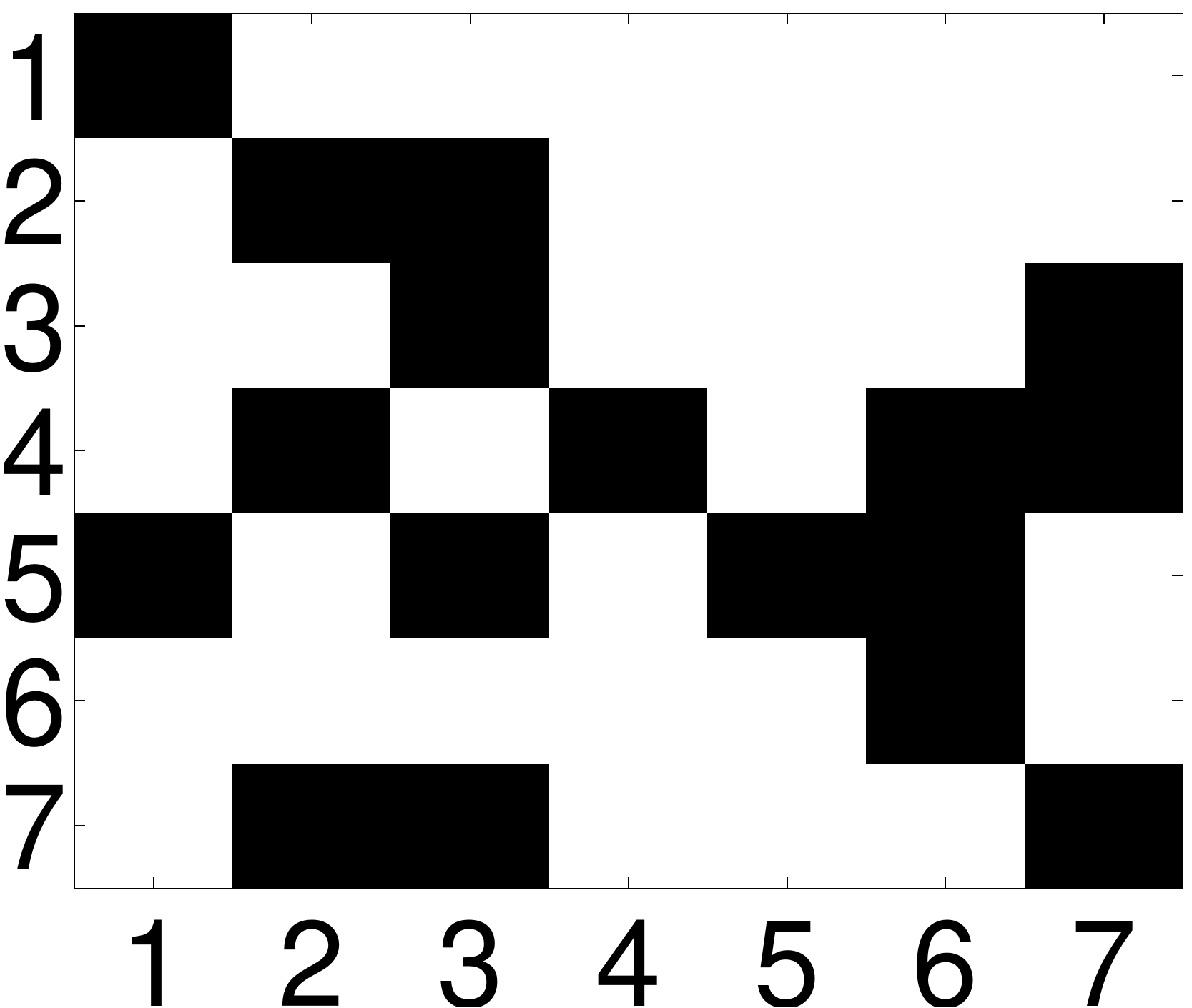} &
\includegraphics[width=0.17\textwidth]{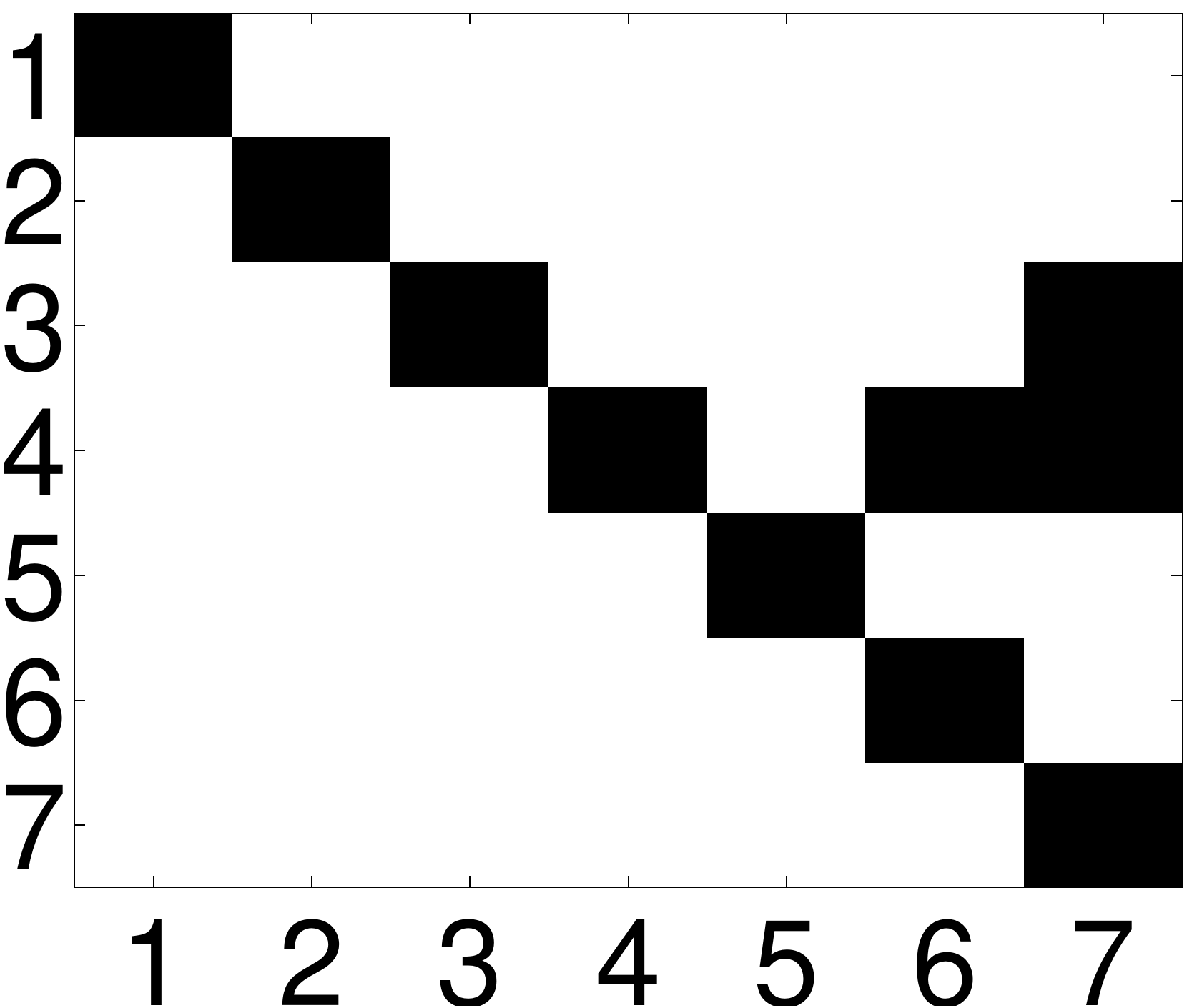} &
\includegraphics[width=0.17\textwidth]{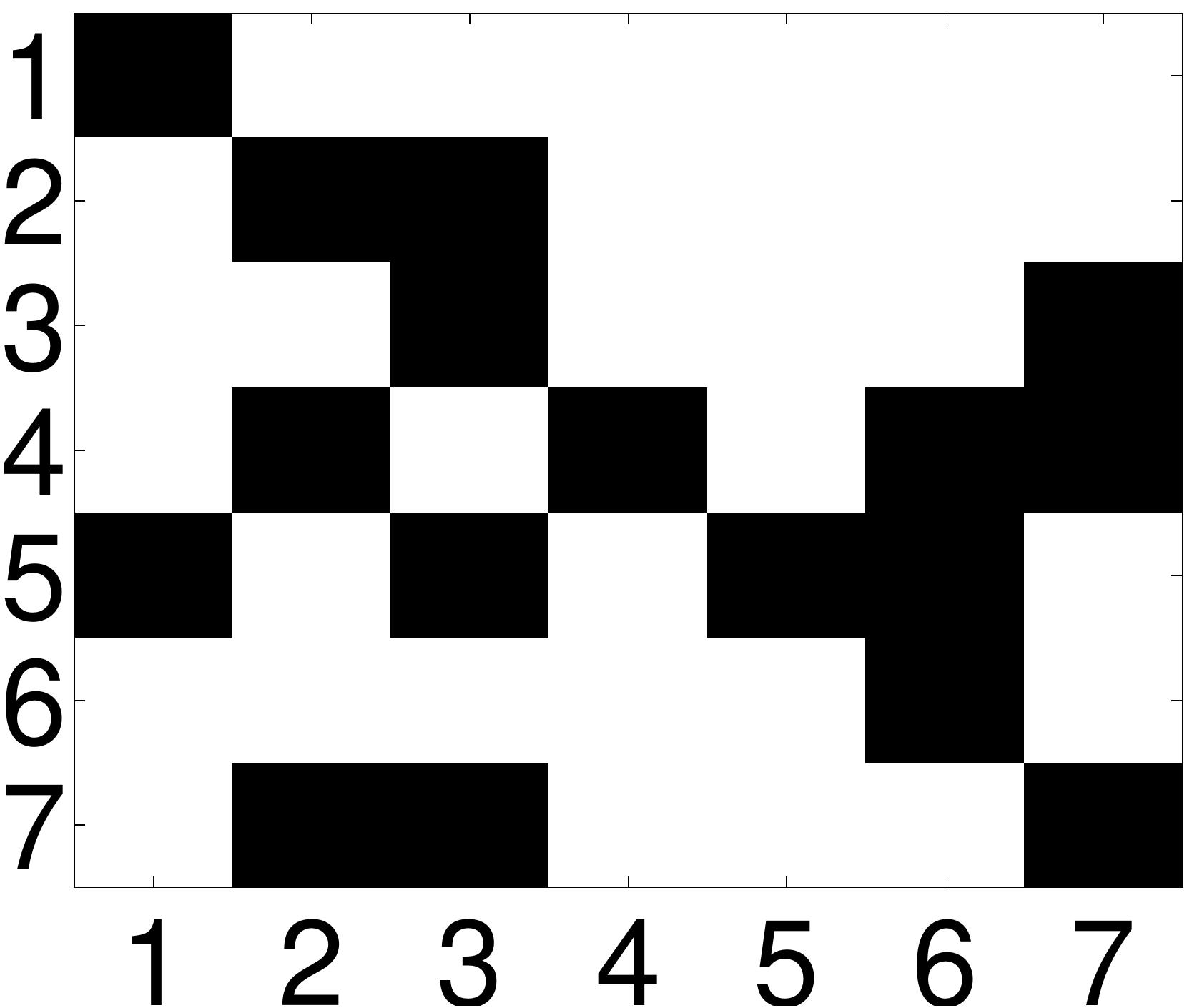} &
\includegraphics[width=0.17\textwidth]{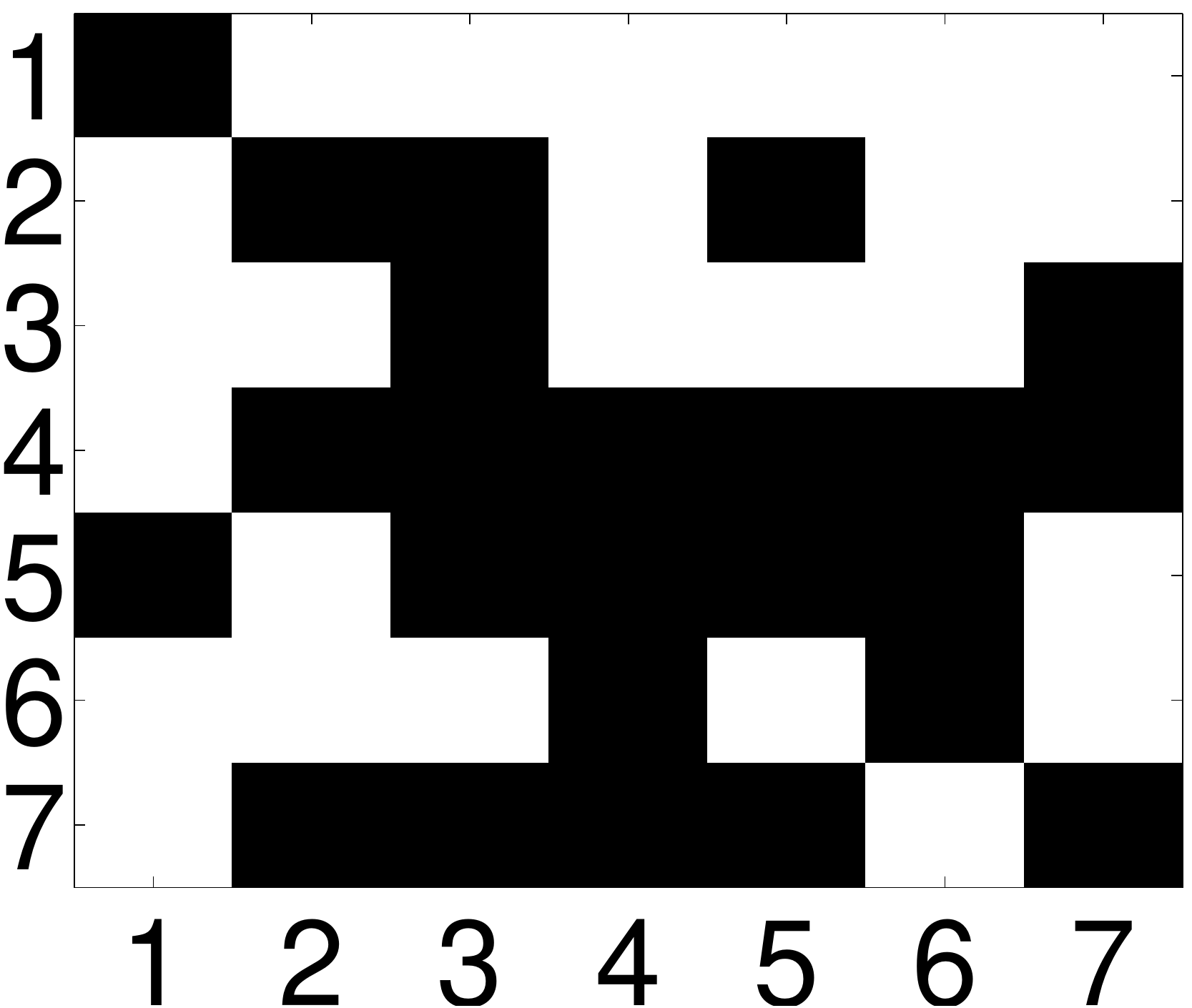} &
\includegraphics[width=0.17\textwidth]{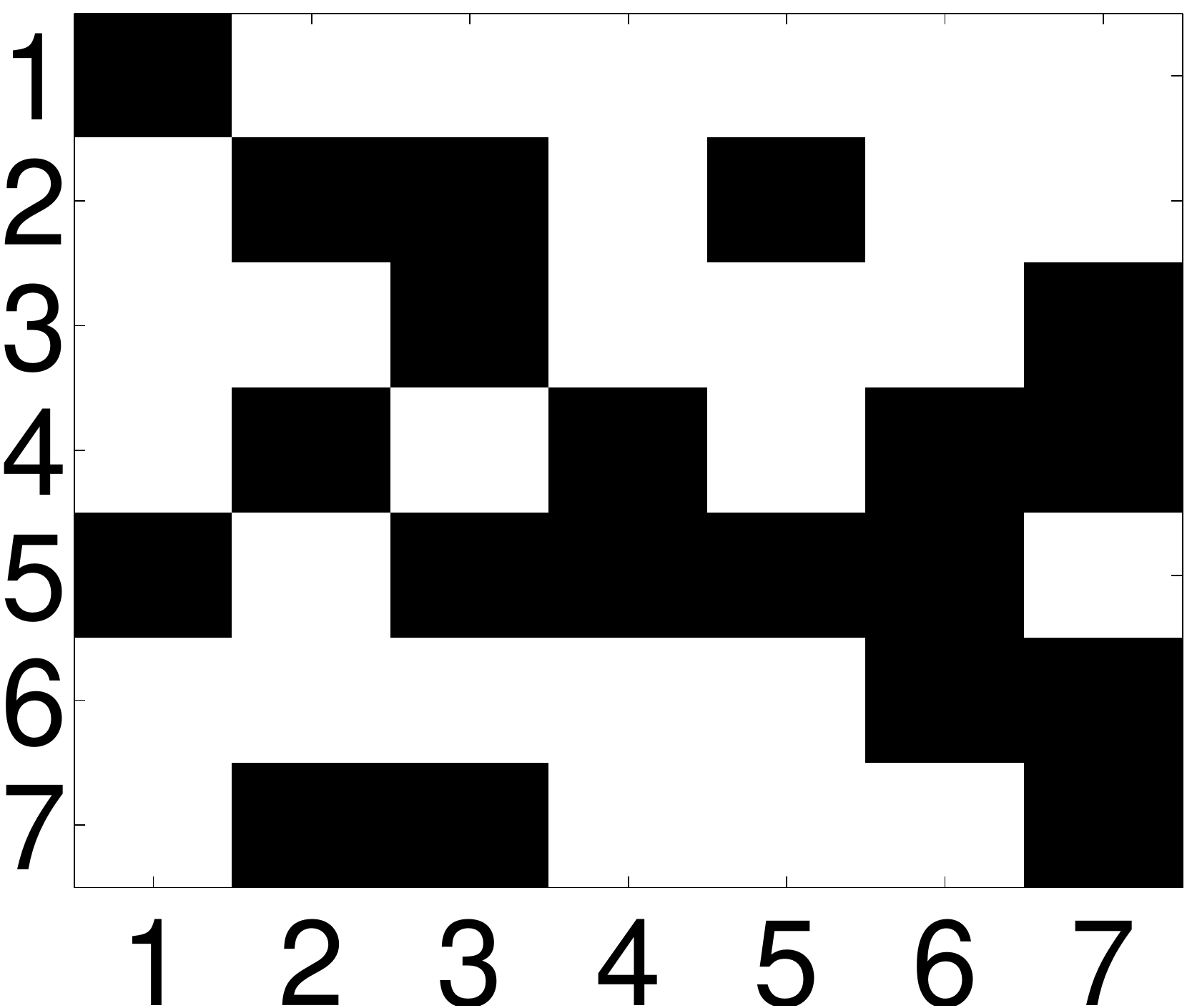} \\
TRUE & GRANGER & RIDGE & LASSO & GLASSO
\end{tabular}
\caption{Simulated causal influence matrix and estimates according to Granger Causality, Ridge Regression, Lasso and Group Lasso.}
\label{fig2}
\end{center}
\end{figure}

First, we illustrate the different behavior of the investigated methods in Figure \ref{fig2}. This example corresponds to the situation without noise and with known model order $P=5$. The left figure shows the true underlying causal structure, with a black box indicating a causal interaction. The reconstructions for the different methods are based on a single estimate of the VAR coefficients. For Granger causality, we use a threshold of 2. For Ridge Regression, we use a significance level of $\gamma=0.05$. For Lasso, Ridge Regression and Group Lasso, the regularizing constant is fixed by using 10-fold cross-validation (with respect to prediction accuracy). We display the binary influence matrix in Figure \ref{fig2}. In this example Ridge Regression exhibits perfect reconstruction and outperforms all other methods. Group Lasso comes second. Note that, due to the strong tendency of Group Lasso to select the same influences for each time lag, its estimated causal dependency matrix is sparser than that of Lasso.

\begin{figure}[htb]
\begin{center}
\begin{tabular}{cccc}
$P=5$ &
\includegraphics[width=0.27\textwidth]{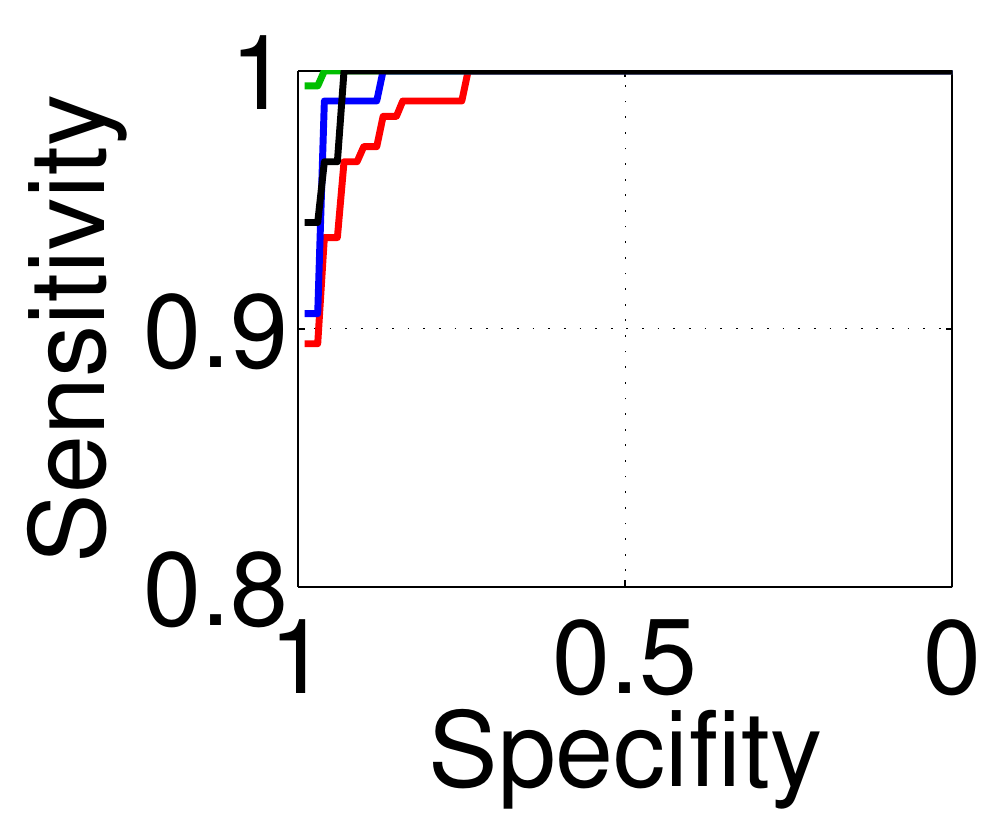} &
\includegraphics[width=0.27\textwidth]{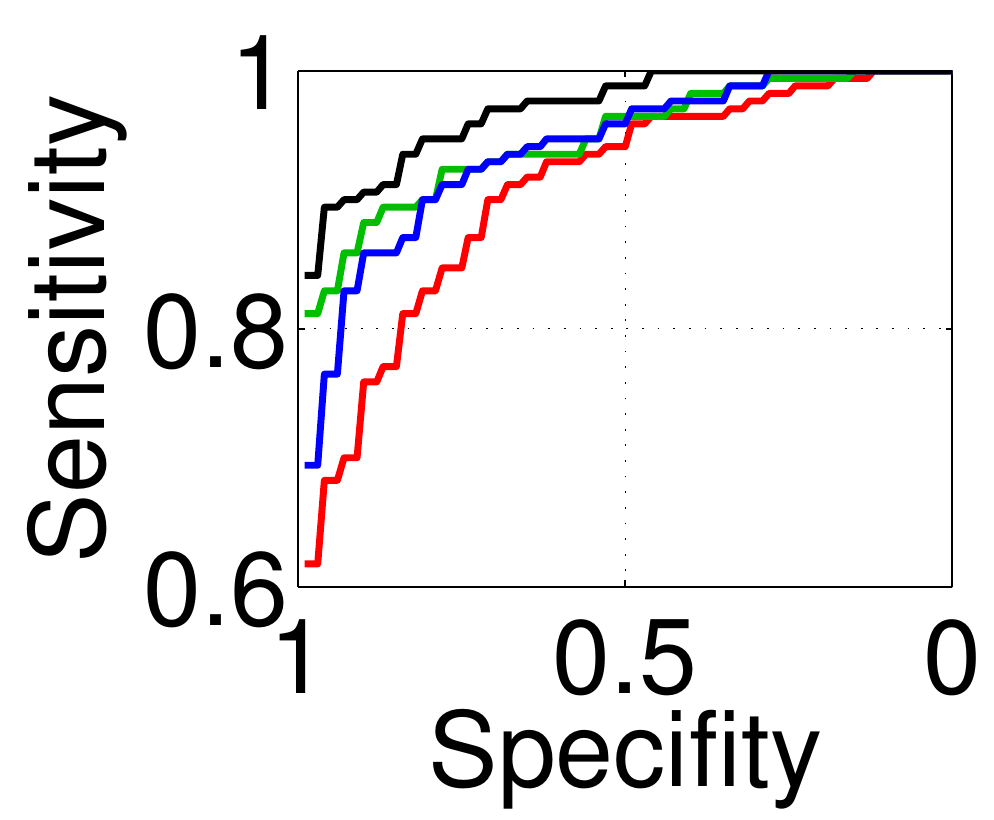} &
\includegraphics[width=0.27\textwidth]{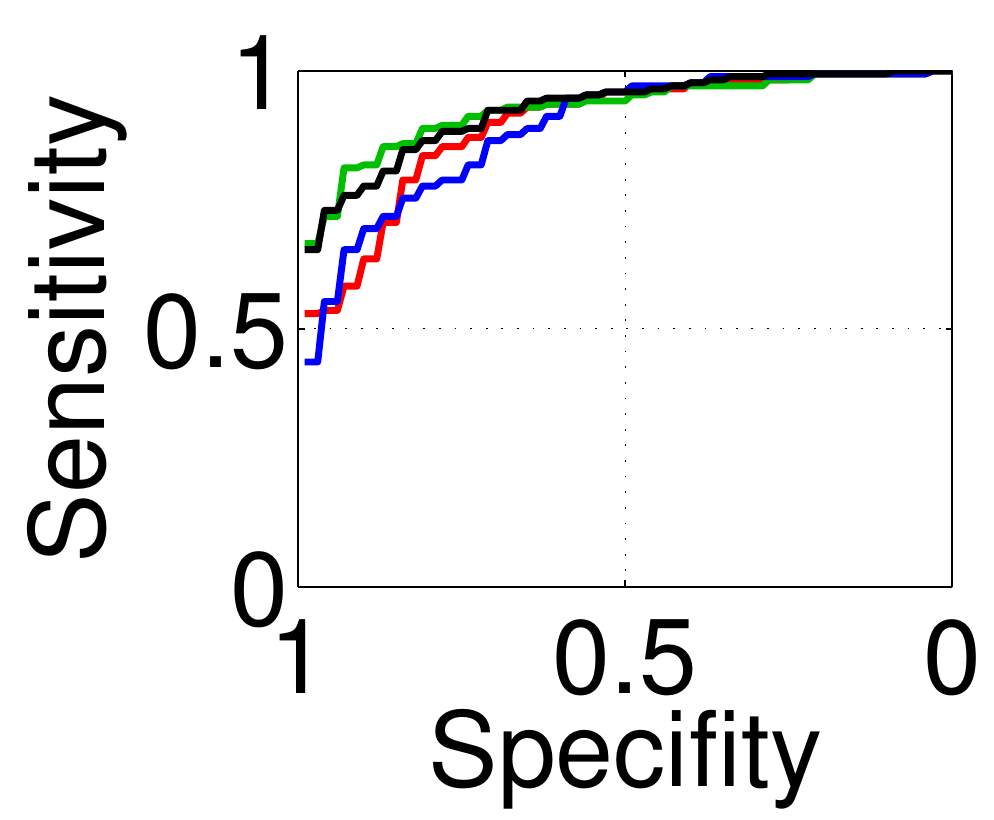} \\
$P=10$ &
\includegraphics[width=0.27\textwidth]{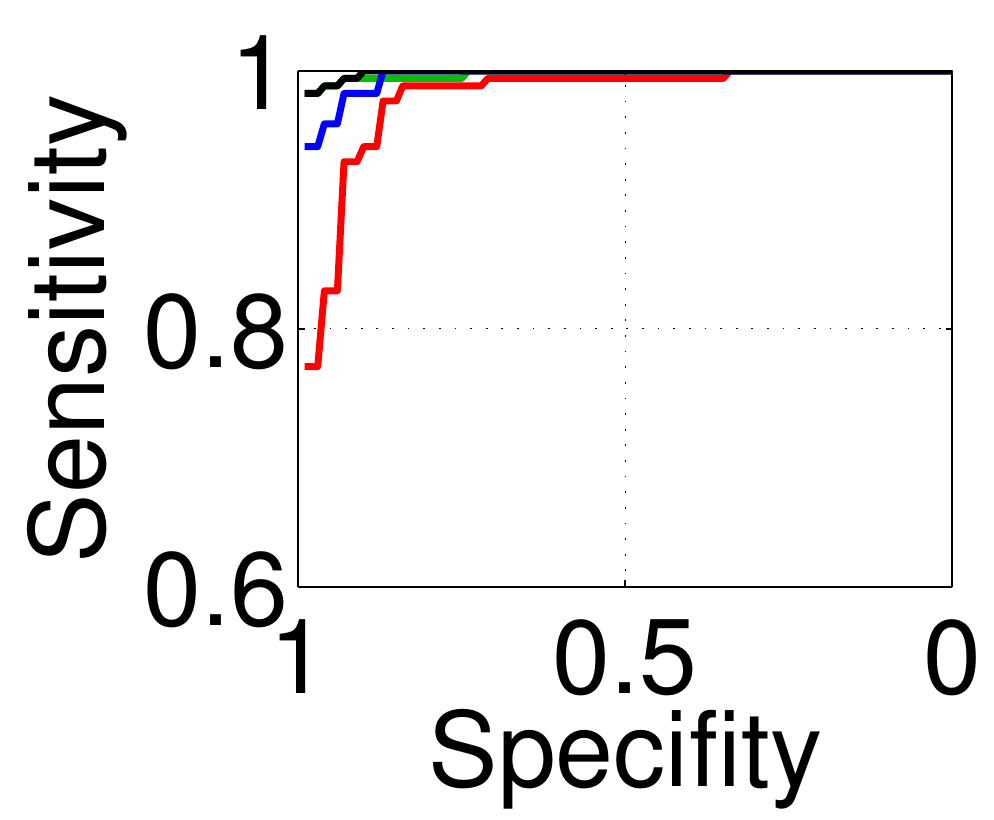} &
\includegraphics[width=0.27\textwidth]{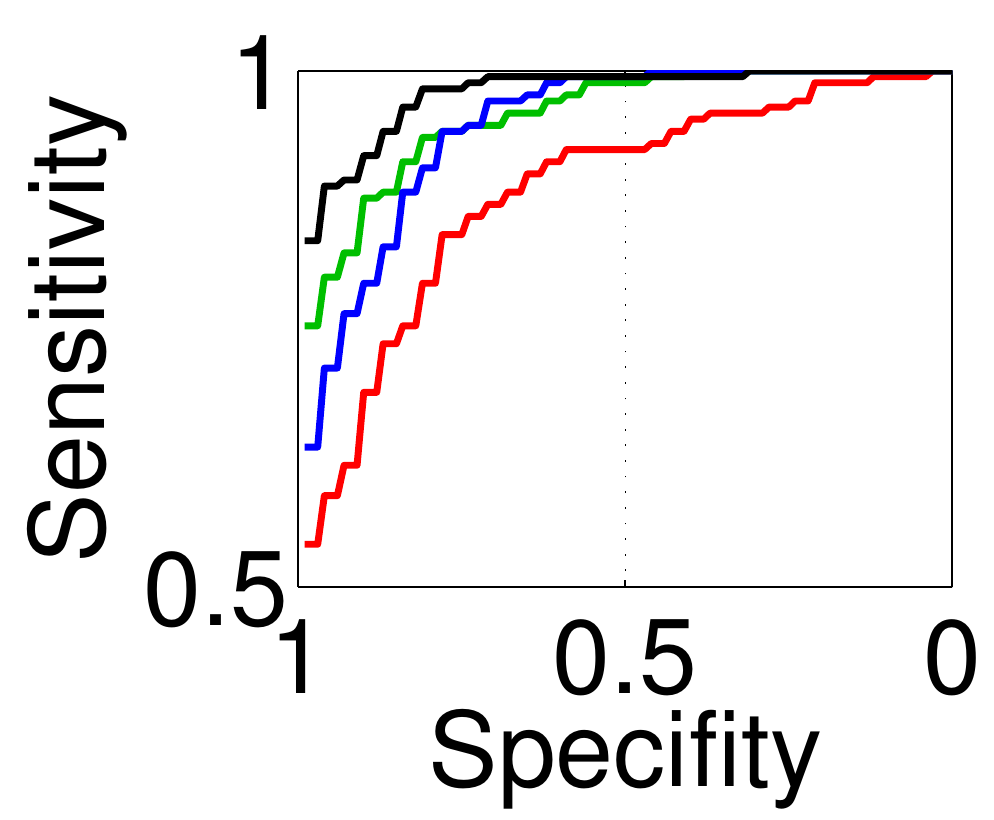} &
\includegraphics[width=0.27\textwidth]{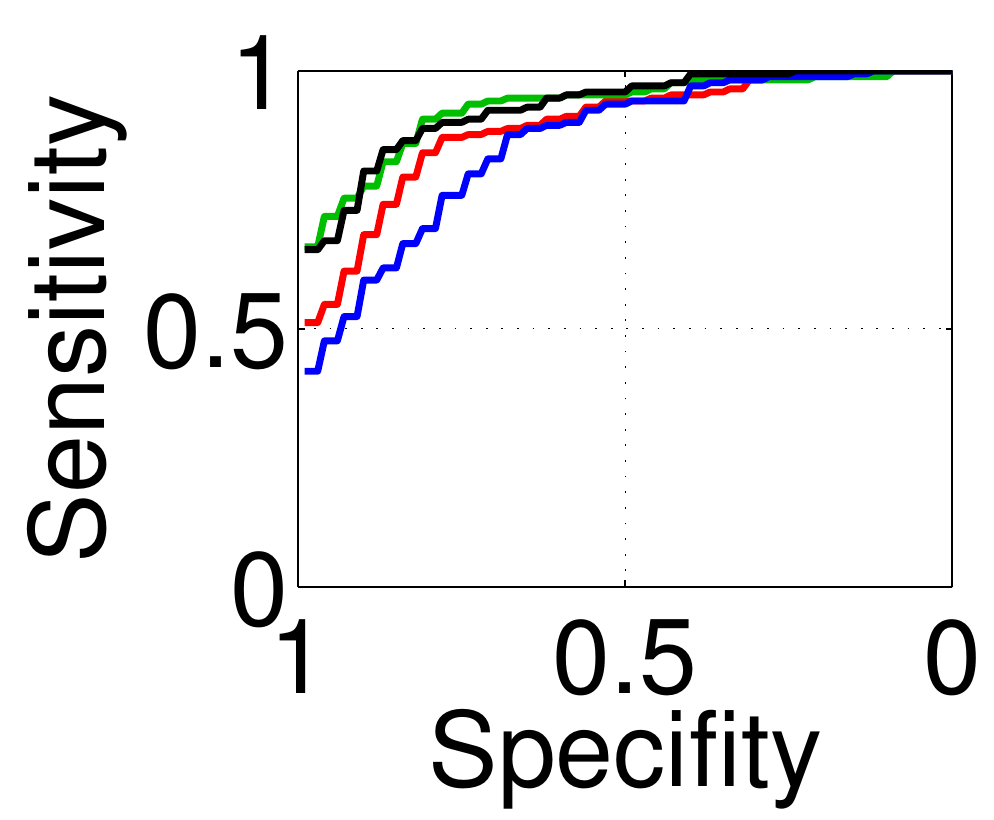} \\
 & NO NOISE & WHITE NOISE & MIXED NOISE
\end{tabular}
\caption{Average ROC curves of Granger Causality (red), Ridge Regression (green), Lasso (blue) and Group Lasso (black) in three different noise conditions and for two different model orders.}
\label{fig1}
\end{center}
\end{figure}

Table~\ref{tab1} summarizes the AUC scores obtained in the experiments described above. The complementing ROC curves are shown in Figure~\ref{fig1}. In short it can be stated that Group Lasso and Ridge Regression outperform their competitors in all scenarios, although not always significantly. While Ridge Regression performs slightly better than Group Lasso in the noiseless condition, Group Lasso has a clearly visible yet insignificant advantage over all methods in the white noise setting. Under the influence of mixed noise Ridge Regression and Group Lasso are on par.  Note furthermore that the ROC curve for Lasso is below the ROC curve of Group Lasso, which shows that Lasso tends to be too dense.  Interestingly, knowledge of the true model order hardly provided any significant advantage in our simulations.

\begin{table}[htb]
\begin{center}
{\small{
\begin{tabular}{clllll}
\hline
 && GRANGER & RIDGE & LASSO & GLASSO \\
\hline
& NO NOISE & 0.991 $\pm$ 0.004 & \textbf{1.000 $\pm$ 0.000} & 0.996 $\pm$ 0.002 & 0.997 $\pm$ 0.002 \\
$P=5$& WHITE NOISE & 0.910 $\pm$ 0.023 & 0.948 $\pm$ 0.020 & 0.941 $\pm$ 0.021 & 0.971 $\pm$ 0.016 \\
& MIXED NOISE & 0.896 $\pm$ 0.012 & \textbf{0.928 $\pm$ 0.010} & 0.889 $\pm$ 0.011 & \textbf{0.926 $\pm$ 0.012} \\
\hline
& NO NOISE & 0.980 $\pm$ 0.005 & \textbf{0.998 $\pm$ 0.002} & \textbf{0.996 $\pm$ 0.002} & \textbf{0.999 $\pm$ 0.001} \\
$P=10$& WHITE NOISE & 0.885 $\pm$ 0.019 & 0.958 $\pm$ 0.012 & 0.948 $\pm$ 0.013 & \textbf{0.979 $\pm$ 0.005} \\
& MIXED NOISE & 0.893 $\pm$ 0.013 & \textbf{0.931 $\pm$ 0.015} & 0.861 $\pm$ 0.014 & \textbf{0.931 $\pm$ 0.007} \\
\hline
\end{tabular}
}}
\caption{Average AUC scores and standard errors of Granger Causality, Ridge Regression, Lasso and Group Lasso in three different noise conditions and for two different model orders. Entries with significant superior score are highlighted.}
\label{tab1}
\end{center}
\end{table}
%-------------------------------------------
\section{Conclusion} \label{sec:conclusion}
%-------------------------------------------
We presented a novel approach for causal discovery in multivariate time series which is based on the Group Lasso. As an alternative we also discussed Ridge Regression with subsequent multiple testing according to \cite{Hothorn0801} which is also novel in the context of VAR modeling. Both approaches were shown to outperform standard methods in simulated scenarios. Future research will aim at applying our techniques to real-world problems. Given that the sparsity assumption is correct, our Group Lasso approach should be able to handle much larger problems than the ones that were considered here by 1) splitting the problem into $M$ independent subproblems and 2) using the active set solver of \cite{Roth0801} in combination with strong regularization that ensures staying in the sparse regime. We expect that this will allow large-scale applications such as the estimation of cerebral information flow from functional Magnetic Resonance Tomography (fMRI) recordings to benefit from the improved accuracy of our approach.

%diagonale bestrafen oder nicht.

% Acknowledgements should go at the end, before appendices and references

\subsection*{Acknowledgements} 
This work was supported in part by the German BMBF (FKZ 01GQ0850, 01-IS07007A and 16SV2234) and the FP7-ICT Programme of the European Community under the PASCAL2 Network of Excellence, ICT-216886. We thank Thorsten Dickhaus for discussions.


\bibliographystyle{apalike}

\end{document}